\documentclass[nofootinbib,showpacs,preprintnumbers,amsmath,amssymb,floatfix]
{revtex4}
\usepackage{epsf}

\begin{document}


\title{Role of the singular factors in the standard  fits for initial  parton  densities}

\vspace*{0.3 cm}

\author{B.I.~Ermolaev}
\affiliation{Ioffe Physico-Technical Institute, 194021
  St.Petersburg, Russia}
\author{M.~Greco}
\affiliation{Department of Physics and INFN, University Rome III,
Rome, Italy}
\author{S.I.~Troyan}
\affiliation{St.Petersburg Institute of Nuclear Physics, 188300
Gatchina, Russia}

\begin{abstract}
Total resummation  of double- and single- logarithms of $x$
contributing to the spin-dependent structure function $g_1$
ensures its steep rise at small $x$. In the asymptotic limit $x
\to 0$, the  resummation  leads to the Regge behavior of $g_1$ and
allows to calculate the non-singlet and singlet intercepts of
$g_1$. DGLAP lacks such a resummation but suggests special
phenomenological fits for the initial parton densities such that
the singular factors $x^{-\alpha}$ in the fits mimic the
resummation and also provide $g_1$ with the steep (power-like)
rise at the small-$x$ region. Accounting for the total resummaton
of logarithms of $x$ allows to drop the singular factors in the
fits and leads to a remarkable simplification of the fits.
\end{abstract}


\maketitle

\section{Introduction}

The standard theoretical instrument for investigating the DIS
structure function $g_1(x, Q^2)$ is DGLAP\cite{dglap}. In this
approach, $g_1^{DGLAP}$ is a convolution of the coefficient
functions $C_{DGLAP}$ and evolved parton distributions which are
also expressed as a convolution of the splitting functions
$P_{DGLAP}$ and initial parton densities. The latter are found
with fitting experimental data at $x$ close $1$ and  $Q^2 \sim
1$~GeV$^2$. There is an obvious asymmetry in treating $Q^2$- and
$x$- contributions in DGLAP. Indeed, the leading $Q^2$-
contributions, $\ln(Q^2)$, are accounted to all orders in
$\alpha_s$ whereas $C_{DGLAP}(x)$ and $P_{DGLAP}(x)$ are known in
first two orders of the perturbative QCD. The reason for such
asymmetry is the fact that originally DGLAP was constructed for
operating at large $Q^2$ and $x$ not so far from 1. In this region
$\ln^k Q^2$ were large whereas $x$- contributions from higher
loops were small and could be neglected. On the contrary, in the
small-$x$ region the situation looks opposite: logarithms of $x$,
namely double logarithmic (DL), i.e. the terms $~(\alpha_s
\ln^2(1/x))^k$, and single logarithms (SL), the terms $~(\alpha_s
\ln(1/x))^k$,with $k
 = 1,2,..$, are becoming quite sizable and should be accounted to all orders
in $\alpha_s$. When the total resummation of DL terms was
done\cite{ber}, it led to new expressions, $g_1^{DL}$, for $g_1$.
In particular, Refs.~\cite{ber} demonstrated that the small-$x$
asymptotics of $g_1^{DL}$ was of the Regge (power-like) form and
was much greater than the well-known small-$x$ asymptotics of
$g_1^{DGLAP}$ (see Eq.~(\ref{dglapas})) at $x \leq 0.01$. However,
this result was strongly criticized (see e.g. Refs.~\cite{blum})
where it was shown that using the standard NLO DGLAP formulae
together with appropriate fits for initial parton densities led to
the opposite result: $g_1^{DL} \ll g_1^{DGLAP}$ at small $x$.
After that the total resummation of logarithms of $x$ for $g_1$
was claimed irrelevant for available range of $x$.

Strictly speaking, comparison of results of Refs.~\cite{ber} and
\cite{blum} could not be done in a straightforward way because
DGLAP -evolution equations have always used the running
$\alpha_s$, with the parametrization
\begin{equation}\label{dglapparam}
\alpha_s^{DGLAP} = \alpha_s(Q^2),
\end{equation}
whereas Refs.~\cite{ber} operated with $\alpha_s$ fixed at an
unknown scale. Trying to set a  scale for $\alpha_s$ when the
approach of Refs.~\cite{ber} is used, Refs.~\cite{kw} suggested
that the argument of $\alpha_s$ should be $Q^2$ like in DGLAP. The
 parametrization $\alpha_s^{DGLAP} = \alpha_s(Q^2)$ appears when
 the argument of $\alpha_s$ in each of the ladder rungs of the involved
 Feynman graphs is $k_{\perp}^2$, with $k$ being the momentum of the upper
 parton (a quark or a gluon) of the rung.
 A deeper investigation of  this matter\cite{egt1}
led us to the conclusion that the DGLAP- parametrization of
Eq.~(\ref{dglapparam}) can be a good approximation at $x$ not far
from $1$ only. Instead of it, a new parametrization was suggested
where the argument of $\alpha_s$ in every ladder rungs is the
virtuality of the horizontal gluon (see Ref.~\cite{egt1} for
detail). This parametrization is universally good for both small
$x$ and large $x$.  It converges to the DGLAP- parametrization at
large $x$ but differs from it at small $x$. Using this new
parametrization allowed us to obtain in Refs.~\cite{egt2}
expressions for $g_1$ accounting for all-order resuummations of DL
and SL terms, including the running $\alpha_s$ effects
\footnote{The parametrization of see Ref.~\cite{egt1} was used
later in Refs.~\cite{kotl} for studying the small-$x$ contribution
to the Bjorken sum rule.}.  In the first place, it was used to
obtain numerical values of the intercepts of the singlet and
non-singlet $g_1$. It is worth to mention that these results were
immediately confirmed\cite{kat} by several independent groups who
fitted available HERMES data and extrapolated them into the
asymptotic region $x \to 0$.

Nevertheless, it is known that, despite DGLAP lacks the total
resummation of $\ln x$, it successfully operates at $x \ll 1$. As
a result, the common opinion was formed that not only the total
resummation of DL contributions in Refs.~\cite{ber} but also the
much more accurate calculations performed in Refs.~\cite{egt2}
should be out of use at available $x$ and might be of some
importance in a distant future at extremely small $x$. In
Ref.~\cite{egt3} we argued against such a point of view and
explained why DGLAP can be so successful at small $x$: in order to
be able to describe the available experimental data, DGLAP uses
the singular fits (see for example Refs.~\cite{a,v}) for the
initial parton densities. Singular factors (i.e. the factors which
$\to \infty$ when $x \to 0$ ) in the fits mimic the total
resummaton of Refs.~\cite{egt2}. Using the results of
Ref~\cite{egt2} allows to simplify the quite complicated structure
of the standard DGLAP fits down to normalization constants at
small $x$.

\section{Difference between DGLAP and our approach}

 In DGLAP, $g_1$ is expressed through convolutions of the coefficient functions
and evolved parton distributions. As convolutions look simpler in
terms of integral transforms, it is convenient to represent $g_1$
in the form of the Mellin integral. For example, the non-singlet
component  of $g_1$ can be represented as follows:

\begin{equation}
\label{fdglapmellin} g^{NS}_{1~DGLAP}(x, Q^2) = (e^2_q/2)
\int_{-\imath \infty}^{\imath \infty} \frac{d \omega}{2\imath
\pi}(1/x)^{\omega} C_{DGLAP}(\omega) \delta q(\omega) \exp \Big[
\int_{\mu^2}^{Q^2} \frac{d k^2_{\perp}}{k^2_{\perp}}
\gamma_{DGLAP}(\omega, \alpha_s(k^2_{\perp}))\Big]
\end{equation}
with $C_{DGLAP}(\omega)$ being the non-singlet coefficient
functions, $\gamma_{DGLAP}(\omega,  \alpha_s)$ the non-singlet
anomalous dimensions and $\delta q(\omega)$ the initial
non-singlet quark densities in the  Mellin (momentum) space. The
expression for the singlet $g_1$ is similar, though more involved.
Both $\gamma_{DGLAP}$ and $C_{DGLAP}$  are known in first two
orders of the perturbative QCD. Technically, it is simpler to
calculate them at integer values of $\omega = n$. In this case,
the
 integrand of Eq.~(\ref{fdglapmellin})  is called the $n$-th momentum of
 $g^{NS}_1$. When the moments for different $n$ are known, $g^{NS}$ at arbitrary values of $\omega$
 is obtained with interpolation of the moments.
 Expressions for the initial quark densities are defined from phenomenological consideration,
 with fitting experimental data at $x \sim 1$.  Eq.~(\ref{fdglapmellin})  shows that $\gamma_{DGLAP}$
 govern the $Q^2$- evolution whereas  $C_{DGLAP}$ evolve
 $\delta q(\omega)$
 in the $x$-space from $x \sim 1$ into the small$x$ region.
 When, at the $x$-space, the initial parton distributions $\delta
q(x)$ are regular in $x$, i.e. do not $\to \infty$ when $x \to 0$,
the small-$x$ asymptotics of $g_{1~DGLAP}$ is given by the
well-known expression:
\begin{equation}\label{dglapas}
 g^{NS}_{1~DGLAP},~~g^{S}_{1~DGLAP} \sim \exp \Big[ \sqrt{\ln (1/x) \ln
 \Big( \ln(Q^2/\mu^2)/\ln (\mu^2/
 \Lambda^2_{QCD})\Big)}~\Big].
\end{equation} On the contrary, when the total resummation of the double-logarithms (DL) and single-
logarithms of $x$ is done\cite{egt1}, the Mellin representation
for $g_1^{NS}$  is
\begin{equation}
\label{gnsint} g_1^{NS}(x, Q^2) = (e^2_q/2) \int_{-\imath
\infty}^{\imath \infty} \frac{d \omega}{2\pi\imath }(1/x)^{\omega}
C_{NS}(\omega) \delta q(\omega) \exp\big( H_{NS}(\omega)
\ln(Q^2/\mu^2)\big)~,
\end{equation}
with new coefficient functions  $C_{NS}$,
\begin{equation}
\label{cns} C_{NS}(\omega) =\frac{\omega}{\omega - H_{NS}(\omega)}
\end{equation}
and anomalous dimensions $H_{NS}$,
\begin{equation}
\label{hns} H_{NS} = (1/2) \Big[\omega - \sqrt{\omega^2 -
B(\omega)} \Big]
\end{equation}
where
\begin{equation}
\label{b} B(\omega) = (4\pi C_F (1 +  \omega/2) A(\omega) +
D(\omega))/ (2 \pi^2)~.
\end{equation}
 $ D(\omega)$ and $A(\omega)$ in Eq.~(\ref{b}) are
expressed in terms of  $\rho = \ln(1/x)$, $\eta =
\ln(\mu^2/\Lambda^2_{QCD})$, $b = (33 - 2n_f)/12\pi$ and the color
factors
 $C_F = 4/3$, $N = 3$:

\begin{equation}
\label{d} D(\omega) = \frac{2C_F}{b^2 N} \int_0^{\infty} d \rho
e^{-\omega \rho} \ln \big( \frac{\rho + \eta}{\eta}\big) \Big[
\frac{\rho + \eta}{(\rho + \eta)^2 + \pi^2} + \frac{1}{\rho +
\eta}\Big] ~,
\end{equation}

\begin{equation}
\label{a} A(\omega) = \frac{1}{b} \Big[\frac{\eta}{\eta^2 + \pi^2}
- \int_0^{\infty} \frac{d \rho e^{-\omega \rho}}{(\rho + \eta)^2 +
\pi^2} \Big].
\end{equation}
$H_{S}$  and $C_{NS}$ account for DL and SL contributions to all
orders in $\alpha_s$. When $x \to 0$,
\begin{equation}
\label{gnsas}g_1^{NS} \sim \big( x^2/Q^2\big)^{\Delta_{NS}/2},~
g_1^{S} \sim \big( x^2/Q^2\big)^{\Delta_{S}/2}
\end{equation}
where the non-singlet and singlet intercepts are $\Delta_{NS} =
0.42,~\Delta_{S} = 0.86$. The $x$- behavior of Eq.~(\ref{gnsas})
is much steeper than the one of Eq.~(\ref{dglapas}). Obviously,
the total resummation of logarithms of $x$ leads to the faster
growth of $g_1$ when $x$ decreasing compared to the one predicted
by DGLAP, providing the input $\delta q$ in
Eq.~(\ref{fdglapmellin}) is a regular function of $\omega$ at
$\omega \to 0$.

\section{Structure of the standard DGLAP fits}
Although there are different fits for  $\delta q(x)$ in
literature, all available fits include  both  regular and singular
factors when $x \to 0$ (see e.g. Refs.~\cite{a,v} for detail). For
example, one of the fits from Ref.~\cite{a} is
\begin{equation}
\label{fita} \delta q(x) = N \eta x^{- \alpha} \Big[(1
-x)^{\beta}(1 + \gamma x^{\delta})\Big],
\end{equation}
with $N,~\eta$ being a normalization, $\alpha = 0.576$, $\beta =
2.67$, $\gamma = 34.36$ and $\delta = 0.75$. In the $\omega$
-space Eq.~(\ref{fita}) is a sum of pole contributions:
\begin{equation}
\label{fitaomega} \delta q(\omega) = N \eta \Big[ (\omega -
\alpha)^{-1} + \sum m_k (\omega + \lambda_k)^{-1}\Big],
\end{equation}
with $\lambda_{k} > 0$, so that the first term in
Eq.~(\ref{fitaomega}) corresponds to the singular factor
$x^{-\alpha}$ of Eq.~(\ref{fita}). When the fit Eq.~(\ref{fita})
is  substituted in Eq.~(\ref{fdglapmellin}),  the singular factor
$x^{-\alpha}$ affects the small -$x$ behavior of $g_1$ and changes
its asymptotics Eq.~(\ref{dglapas}) for $g_1$ for the Regge
asymptotics. Indeed, the small- $x$ asymptotics is governed by the
leading singularity $\omega = \alpha$, so
\begin{equation}\label{asdglap}
g_{1~DGLAP} \sim C(\alpha)(1/x)^{\alpha}\Big((\ln(Q^2/\Lambda^2))/
(\ln(\mu^2/\Lambda^2))\Big)^{\gamma(\alpha)}.
\end{equation}
 Obviously,  the actual DGLAP asympotics of Eq.~(\ref{asdglap}) is of the Regge type,
 it differs a lot from the conventional DGLAP asympotics of
 Eq.~(\ref{dglapas}) and
 looks similar to
 our asymptotics given by Eq.~(\ref{gnsas}):  incorporating the singular
factors into DGLAP fits ensures the steep rise of $g_1^{DGLAP}$ at
small $x$ and thereby leads to the success of DGLAP at small $x$.
Ref.~\cite{egt3} demonstrates that without the singular factor
$x^{-\alpha}$ in the fit of Eq.~(\ref{fita}), DGLAP would not be
able to operate successfully at $x \leq 0.05$. In other words, the
singular factors in DGLAP fits
 mimic the total resummation of logarithms of $x$ of
Eqs.~(\ref{gnsint},\ref{gnsas}).  Although both (\ref{asdglap})
and (\ref{gnsas}) predict the Regge asymptotics for $g_1$, there
is a certain difference between them: Eq.~(\ref{asdglap}) predicts
that the intercept of $g_1^{NS}$ should be $\alpha = 0.57$.  As
$\alpha$ is greater than the non-singlet intercept $ \Delta_{NS} =
0.42$, the non-singlet $g_1^{DGLAP}$ grows, when $x \to 0$, faster
than our predictions. However, such a rise is too steep. It
contradicts the results obtained in Refs.~\cite{egt2} and
confirmed in Refs.~\cite{kat}. Usually, the DGLAP equations for
the non-singlets are written in the $x$-space as convolutions of
splitting functions $P_{qq}$ with evolved parton distributions
$\Delta q$ and  the latter are written as another convolution:

\begin{equation}\label{evol}
  \Delta q(x)  = C_q(x,y)\otimes \delta q(y) ,
\end{equation}
with $C_q$ being the coefficient function. Written in this way,
$\Delta q$ is sometimes  believed to be less singular than $\delta
q$ because of the evolution. However applying the Mellin transform
to Eq.~(\ref{evol}) immediately disproves it.

\section{Conclusion}

Comparison of Eqs.~(\ref{dglapas}) and (\ref{asdglap}) shows
explicitly that the singular factor $x^{-\alpha}$ in the
Eq.~(\ref{fita}) for the initial quark density converts the
exponential DGLAP-asympotics into the Regge one. On the other
hand, comparison of Eqs.~(\ref{gnsas}) and (\ref{asdglap})
demonstrates that the singular factors in the DGLAP fits mimic the
total resummation of logarithms of $x$. These factors can be
dropped when the total resummation of logarithms of $x$ performed
in Ref.~\cite{egt2} is taken into account. The remaining, regular
$x$-terms of the DGLAP fits (the terms in squared brackets in
Eq.~(\ref{fita})) can obviously be simplified or even dropped at
small $x$ so that the rather complicated DGLAP fits can be
replaced by constants. It immediately leads to an interesting
conclusion: the DGLAP fits for $\delta q$ have been commonly
believed to represent non-perturbative QCD effects but they
actually mimic the contributions of the perturbative QCD, so the
whole impact of the non-perturbative QCD on $g_1$ at small $x$ is
not large and can be approximated by normalization constants.

\section{Acknowledgement}
B.I.~Ermolaev is grateful to the Organizing committee of the
workshop Spin-05 for financial support of his participation in the
workshop.


\begin{thebibliography}{99}

\bibitem{dglap} G.~Altarelli and G.~Parisi, Nucl.~Phys.B126 (1977) 297;
V.N.~Gribov and L.N.~Lipatov, Sov.~J.~Nucl.~Phys. 15 (1972) 438;
L.N.Lipatov, Sov.~J.~Nucl.~Phys. 20 (1972) 95; Yu.L.~Dokshitzer,
Sov.~Phys.~JETP 46 (1977) 641.

\bibitem{ber} B.I.~Ermolaev, S.I.~Manaenkov and M.G.~Ryskin.
Z.Phys.C69(1996)259; ~J.~Bartels, B.I.~Ermolaev and M.G.~Ryskin.
Z.Phys.C 70(1996)273; J.~Bartels, B.I.~Ermolaev and M.G.~Ryskin.
Z.Phys.C 72(1996)627.

\bibitem{blum} J.~Blumlein, A.~Vogt. Acta Phys. Polon. B27 (1996)
1309. J.~Blumlein, S.~Riemersma, A.~Vogt. Nucl.Phys.Proc.Suppl.51C
(1996) 30; Acta Phys. Poloon. B28 (1997) 577.

\bibitem{kw} J.~Kwiecinski and B.~Ziaja. Phys. Rev.
D60(1999)054004; J.~Kwiecinski, M.~Maul. Phys.Rev.D67(2003)03401;
B.~Ziaja. Phys.Rev.D66(2002)114017; A.~Kotlorz, D.~Kotlorz. Acta
Phys.Polon. B34(2003)2943; Acta Phys.Polon. B35(2004)705.

\bibitem{egt1} B.I.~Ermolaev, M.~Greco and S.I.~Troyan.  Phys.Lett.B
522(2001)57.

\bibitem{kotl} D.~Kotlorz A.~Kotlorz.
Acta Phys. Polon. B35(2004)2503; hep-ph/0407040.

\bibitem{egt2} B.I.~Ermolaev, M.~Greco and S.I.~Troyan.
 Nucl.Phys.B 594 (2001)71; ibid 571(2000)137;
Phys.Lett.B579(321),2004.

\bibitem{kat} J.~Soffer and O.V.~Teryaev. Phys. Rev.56( 1997)1549;
A.L.~Kataev, G.~Parente, A.V.~Sidorov. Phys.Part.Nucl 34(2003)20;
Nucl.Phys.A666(2000)184; A.V.~Kotikov, A.V.~Lipatov, G.~Parente,
N.P.~Zotov. Eur.Phys.J.C26(2002)51; V.G.~Krivohijine,
A.V.~Kotikov, hep-ph/0108224; A.V.~Kotikov, D.V.~Peshekhonov
hep-ph/0110229; N.I.~Kochelev, K.~Lipka, W.D.~Novak,
A.V.~Vinnnikov. Phys. Rev. D67(2003) 074014.

\bibitem{egt3} B.I.~Ermolaev, M.~Greco and S.I.~Troyan. Phys.
Lett. B622 (2005) 93 (hep-ph/0503019).

\bibitem{a} G.~Altarelli, R.D.~Ball, S.~Forte and G.~Ridolfi,
Nucl.~Phys.~B496 (1997) 337; Acta Phys. Polon. B29(1998)1145;

\bibitem{v}A.~Vogt. hep-ph/0408244.




\end{thebibliography}
\end{document}